\documentclass{ws-ijmpa}
\usepackage[super,compress]{cite}
\usepackage{graphicx}
\usepackage{amsmath}
\usepackage{bm}

\DeclareMathOperator{\tr}{tr}

\begin{document}
\markboth{Milton, Pourtolami, Kennedy}{Quantum Self-Propulsion and Torque
of an Inhomogeneous Object out of Thermal Equilibrium}

%
\catchline{}{}{}{}{}
%

\title{Quantum Vacuum Self-Propulsion and Torque}

\author{Kimball A. Milton
}

\address{H.L. Dodge Department of Physics and Astronomy,\\
Norman, OK 73019 USA\\
kmilton@ou.edu}

\author{Nima Pourtolami}

\address{National Bank of Canada\\
Montreal, Quebec H3B 4S9 Canada\\
nima.pourtolami@gmail.com}

\author{Gerard Kennedy}
\address{School of Mathematical Sciences, University of Southampton\\
Southampton SO17 1BJ UK\\
g.kennedy@soton.ac.uk}

\maketitle

\begin{history}
\received{Day Month Year}
\revised{Day Month Year}
\end{history}

\begin{abstract}
This article summarizes our recent efforts
to understand spontaneous quantum vacuum
forces and torques, which require that a stationary 
 object be out of thermal equilibrium with
the blackbody background radiation.  We proceed by a systematic expansion in
powers of the electric susceptibility.  In first order, no spontaneous
force can arise, although a torque can appear, but only if the body is composed
of nonreciprocal material.  In second order, both forces and torques can 
appear, with ordinary materials, but only if the body is inhomogeneous.
In higher orders, this last requirement may be removed.  We give a number of 
examples of bodies displaying  second-order spontaneous  forces and torques, 
some of which might be amenable to observation.

\end{abstract}

\ccode{PACS numbers: 42.50.Lc, 05.70.Ln, 68.35.Af}


\section{Introduction}	

Growing out of our explorations of quantum (Casimir) 
friction,\cite{eqfi,eqfii,qfpc} we
  are studying the forces and torques arising on bodies in vacuum out
  of thermal equilibrium.
 A torque can arise in first order in susceptibility for a body that
 is nonreciprocal. This is described in more detail in 
Refs.~\citen{nr,kennedy}.
 For a reciprocal body, spontaneous torques and forces require not
 only nonequilibrium but inhomogeneity of the body.  Such effects emerge in 
second order.  More detail on spontaneous forces and torques can be found in 
Refs.~\citen{selfprop, sptorque}.  See these papers for fuller references to
the literature.

 We use natural units, $\hbar=c=\epsilon_0=\mu_0=k_B=1$.



\section{Quantum Vacuum Torque in Nonreciprocal Media}
\label{sec2}
This phenomenon was first discussed in Ref.~\citen{Strekha}, and then
subsequently in our paper, Ref.~\citen{nr}.  We summarize the latter here.

Classically, the torque on a stationary dielectric
body with polarization vector $\mathbf{P}$ is
given by\cite{CE}
\begin{equation}
\bm{\tau}=\int (d\mathbf{r})\frac{d\omega}{2\pi}\frac{d\nu}{2\pi}
e^{-i(\omega+\nu)t}\left[\mathbf{P(r};\omega)\times \mathbf{E(r};\nu)+
P_i(\mathbf{r};\omega)(\mathbf{r}\times\bm{\nabla})E_i(\mathbf{r};\nu)
\right].
\end{equation}
The first term here is called the {\it internal\/} torque and the second the
{\it external\/} torque, 
because the latter is reflective of the force on the body.
We expand this out to first order in generalized susceptibilities, using
\begin{equation}
\mathbf{E}^{(1)}(\mathbf{r};\omega)=\int (d\mathbf{r'})\bm{\Gamma}
(\mathbf{r-r'};\omega)\cdot
\mathbf{P}(\mathbf{r}';\omega),\quad
\mathbf{P}^{(1)}(\mathbf{r};\omega)=\bm{\chi}(\mathbf{r};\omega)
\cdot\mathbf{E}(\mathbf{r};\omega),
\end{equation}
in terms of the (local) electric susceptibility $\bm{\chi}$ and
the (vacuum) electromagnetic Green's dyadic $\bm{\Gamma}$.
We evaluate the two contributions (PP and EE)
to the quantum vacuum torque by use of the 
fluctuation-dissipation theorem (FDT):
\begin{subequations}
\begin{eqnarray}
\langle \mathbf{P}_i(\mathbf{r};\omega)\mathbf{P}_j(\mathbf{r}';\nu)\rangle&=&
2\pi\delta(\omega+\nu)\delta(\mathbf{r-r'})\bm{\chi}^A_{ij}(\mathbf{r};\omega)\coth
\frac{\beta'\omega}2,
\\
\langle\mathbf{E}_i(\mathbf{r};\omega)\mathbf{E}_j(\mathbf{r}';\nu)\rangle&=&
2\pi\delta(\omega+\nu)\Im\bm{\Gamma}_{ij}(\mathbf{r-r'};\omega)\coth
\frac{\beta\omega}2,
\end{eqnarray}
\end{subequations}
in terms of the inverse temperatures 
of the vacuum background, $\beta=1/T$, and of the body, $\beta'=1/T'$.

Here, $\bm{\Gamma}$ is taken to be the usual vacuum retarded Green's dyadic, 
which at coincident points is (rotationally averaged)
\begin{equation}
    \bm{\Gamma}(\mathbf{r-r'};\omega)
\to\bm{1}\left(\frac{\omega^2}{6\pi R}+i\frac{\omega^3}{6\pi}+O(R)\right),
\quad R=|\mathbf{r-r'}|\to0,
\end{equation}
while $\bm{\chi}^A$ is the anti-Hermitian part of the susceptibility:
\begin{equation}
    \bm{\chi}^A=\frac1{2i}(\bm{\chi}-\bm{\chi}^\dagger),
 \quad   \Im\bm{\chi}^A_{ij}(\omega)=
-\frac12\Re[\chi_{ij}(\omega)-\chi_{ji}(\omega)],
\end{equation} which displays the antisymmetric part of $\bm{\chi}^{A}$, 
which is  even in $\omega$.  The real part of $\chi^A$ is symmetric in 
indices and odd in $\omega$. Only the former contributes to the torque.

Using this, we readily calculate the torque on a nonreciprocal body in vacuum,
due to PP and EE fluctuations:
\begin{equation}
    \tau_i=-\int_{-\infty}^\infty \frac{d\omega}{2\pi} \frac{\omega^3}{6\pi}
\left[\coth\frac{\beta'\omega}2-\coth\frac{\beta\omega}2\right]\epsilon_{ijk}
\Re\alpha_{jk}(\omega),
    \end{equation}
    where the mean polarizability of the body is given by
 $\alpha_{jk}(\omega)=\int(d\mathbf{r})\chi_{jk}(r;\omega)$.
This result exactly agrees with that of Strekha et al.\cite{Strekha}  
However, there is no quantum
vacuum force in this static situation.
To create a nonreciprocal medium typically requires an external magnetic
field, so this is not exactly a pure vacuum phenomenon.
If the nonreciprocity is generated by a magnetic field of 1 T, the 
corresponding torque for a gold nanoball of radius 100 nm
 is $\sim 10^{-24}$ N\,m for $T'=600$~K, $T=300$ K.

\section{Self-Propulsive Force}

The Lorentz force on a dielectric body is\cite{CE}, where we have
omitted a term which vanishes when the FDT is used, 
\begin{equation}
\mathbf{F}=-\int(d\mathbf{r})\int \frac{d\omega}{2\pi}\frac{d\nu}{2\pi}
e^{-i(\omega+\nu)t}
\frac\omega\nu
P_i(\mathbf{r};\omega)\cdot(\bm{\nabla})\cdot
E_i(\mathbf{r};\nu).\label{Lorentzforce}
\end{equation}
Now we expand the fields out to second order (4th order in generalized
susceptibilities, $\bm{\chi}$ and $\bm{\Gamma}$, after use of the FDT):
\begin{subequations}
\begin{eqnarray}
\mathbf{E}^{(2)}(\mathbf{r};\nu)
&=&\int(d\mathbf{r'})\,\bm{\Gamma}(\mathbf{r-r'};\nu)
\cdot\bm{\chi}(\mathbf{r}';\nu)\cdot\mathbf{E}(\mathbf{r'};\nu),\\
\mathbf{E}^{(3)}(\mathbf{r};\nu)
&=&\int(d\mathbf{r'})(d\mathbf{r''})\,\bm{\Gamma}(\mathbf{r-r'};\nu)
\cdot\bm{\chi}(\mathbf{r}';\nu)\cdot\bm{\Gamma}(\mathbf{r'-r''};\nu)\cdot
\mathbf{P(r'';\nu)},\\
\mathbf{P}^{(2)}(\mathbf{r};\omega)
&=&\int(d\mathbf{r'})\,\bm{\chi}(\mathbf{r};\omega)
\cdot\bm{\Gamma}(\mathbf{r-r'};\omega)\cdot\mathbf{P}(\mathbf{r'};\omega),\\
\mathbf{P}^{(3)}(\mathbf{r};\omega)
&=&\int(d\mathbf{r'})\,\bm{\chi}(\mathbf{r};\omega)\cdot
\bm{\Gamma}(\mathbf{r-r'};\omega)\cdot\bm{\chi}(\mathbf{r'};\omega)
\cdot\mathbf{E}(\mathbf{r'};\omega).
\end{eqnarray}
\end{subequations}

Using the symmetries of the integrand we find the following general
expression for the force on a body composed of isotropic material:
\begin{eqnarray}
\mathbf{F}&=&\int(d\mathbf{r})(d\mathbf{r'})\int_{-\infty}^\infty 
\frac{d\omega}{2\pi}
X(\mathbf{r,r'};\omega)\Im \Gamma_{ji}(\mathbf{r'-r};\omega)\nonumber\\
&&\qquad\qquad\times\bm{\nabla}\Im \Gamma_{ij}(\mathbf{r-r'};\omega)
\left[\coth\frac{\beta\omega}2-\coth\frac{\beta'\omega}2\right],
\label{qvf}
\end{eqnarray}
where the second-order susceptibility product is
\begin{equation}
X(\mathbf{r,r'};\omega)
=\Im\chi(\mathbf{r};\omega)
\Re\chi(\mathbf{r'};\omega)-\Re\chi(\mathbf{r};\omega)
\Im\chi(\mathbf{r'};\omega).\label{X}
\end{equation}
Obviously, if the body is homogeneous, so the susceptibility is independent
of position, then $X$ vanishes, and there is no force on the body.


Consider  a body composed of two parts, each of
which is separately made of homogeneous material.
In such a case, the force in the $z$ direction  reduces to
\begin{equation}
F_z=8\int_0^\infty
\frac{d\omega}{2\pi}X_{AB}(\omega)I_{AB}(\omega)
\left[\frac1{e^{\beta\omega}-1}-\frac1{e^{\beta'\omega}-1}\right],
\end{equation}
where the geometric integral is
\begin{equation}
I_{AB}(\omega)=
\int_A(d\mathbf{r})\int_B(d\mathbf{r'})
\frac12\nabla_z\left[\Im \Gamma_{ji}(\mathbf{r'-r};\omega)
\Im \Gamma_{ij}(\mathbf{r-r'};\omega)\right],\label{IAB}
\end{equation} and the susceptibility product becomes
\begin{equation}
X_{AB}(\omega)=
\Im\chi_A^{\vphantom{q}}(\omega)
\Re\chi_B^{\vphantom{q}}(\omega)
-\Re\chi^{\vphantom{q}}_A(\omega)
\Im\chi_B^{\vphantom{q}}(\omega).\label{XAB}
\end{equation}

From the form of the vacuum electromagnetic Green's dyadic,
\begin{equation}
    \bm{\Gamma}^{\prime}(\mathbf{r-r'};\omega)=
(\bm{\nabla\nabla}-\bm{1}\nabla^2)\frac{e^{i\omega R}}{4\pi R},\quad 
\bm{\Gamma}^{\prime}=
\bm{\Gamma}+\bm{1},
\end{equation} 
we find for the gradient of the product of the imaginary parts of $\bm{\Gamma}$
\begin{equation}
\frac12\bm{\nabla}\left[\Im\Gamma_{ji}(-\mathbf{R};\omega)
\Im\Gamma_{ij}(\mathbf{R};\omega)\right]
=\frac1{(4\pi)^2}\frac{\mathbf{R}}{R^8}\phi(v),\quad
\mathbf{R}=\mathbf{r-r'},\quad v=\omega R,\label{gradgg}
\end{equation}
where
\begin{equation}
\phi(v)=-9-2v^2-v^4+(9-16v^2+3v^4)\cos2v+v(18-8v^2+v^4)\sin2v.\label{phi}
\end{equation}
The limiting behaviors are
\begin{equation}
\phi(v)\sim -\frac49 v^8+\frac{28}{225}v^{10}+\cdots, v\ll 1,
\phi(v)\sim-v^4+v^5\sin 2v+3v^4 \cos 2v+\cdots,  v\gg1 \label{limits}.
\end{equation}

\begin{figure}
\centering
\begin{minipage}{.4\textwidth}
\centering
\includegraphics[width=5cm]{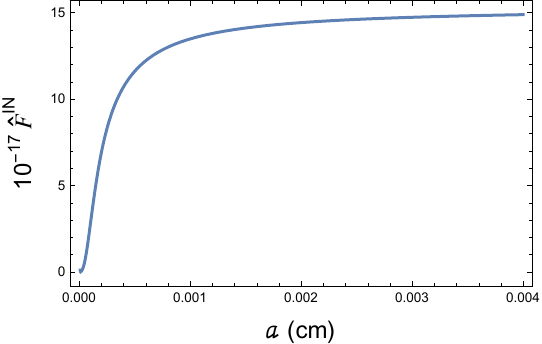}
\caption{Force ($\hat F$)
on inhomogeneous needle for $a=b$ when $T=300$ K and $T'=600$ K.
Note the saturation for large $a$; only the
immediate region near the $A$-$B$ interface contributes.
The saturated force ($F$) is
about $0.03 \chi^{\vphantom{q}}_A$ N.}
\label{fig2}
\end{minipage}\hfill
\begin{minipage}{.4\textwidth}
\centering
\includegraphics[width=5cm]{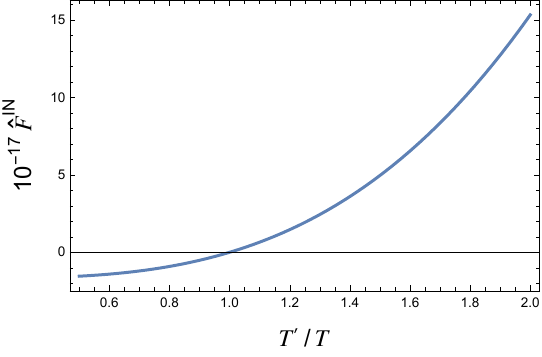}
\caption{Dependence of force ($\hat{F}$) on $T'$ for $T=300$ K.  $a=b=1$ cm.}
\vskip1.2cm

\label{fig3}
\end{minipage}

\end{figure}

{\it Example 1: Thin inhomogeneous needle.}
The susceptibility of a thin needle of cross section $S$ with parts
$A$ and $B$ of length
$a$ and $b$ made of different materials is 
\begin{equation}
\chi(\mathbf{r};\omega)=S\left[\delta(x)\delta(y)\theta(z)\theta(a-z)
\chi_A^{\vphantom{q}}
(\omega)+\delta(x)\delta(y)\theta(-z)\theta(b+z)\chi_B^{\vphantom{q}}(\omega)
\right].
\end{equation}
For definiteness, chose $\chi^{\vphantom{q}}_A$ to be a real constant 
(frequencies small
compared to binding energies), while $B$ is taken to be a Drude-type metal
\begin{equation}
\chi_B^{\vphantom{q}}(\omega)=-\frac{\omega_p^2}{\omega^2+i\omega\nu}.
\label{drude}\end{equation}
For gold, with the vacuum at $T=300$ K, and the cross-sectional radius of the
needle being 1 mm, the force is about ($a_0=1$ cm, $\beta_0=40$ (eV)$^{-1}$)
\begin{equation}
F=-\frac{S^2\omega_p^2\nu\chi^{\vphantom{q}}_A}{120\pi^3}
\frac{\beta_0^2}{a_0^5}\hat{F}=
-2 \chi^{\vphantom{q}}_A \times 10^{-20 }\hat{F}\,\mbox{N},
\end{equation}
where $\hat{F}$ for gold  is shown in Figs.~\ref{fig2} and \ref{fig3}.
Since $\hat F\sim 10^{18}$, this seems very large.

However, this is probably overly optimistic.  For the validity of our
weak-susceptibility model, we need the needle to be much thinner than 1 mm.
The appropriate scale would seem to be the skin depth, which is\cite{CE}
\begin{equation}
\delta=(\omega \sigma/2)^{-1/2}=(\omega^2\Im\chi/2)^{-1/2}
=\sqrt{\frac{2(\omega^2+\nu^2)}{\omega\omega_p^2\nu}}\sim 50 \,\mbox{nm},
\end{equation}
which  reduces this force substantially (by a factor of $10^{-17}$).

\begin{figure}
\centering
\begin{minipage}{.45\textwidth}\centering
\includegraphics[width=5cm]{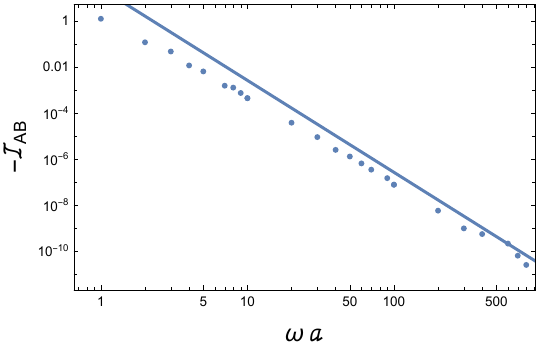}
\caption{Numerical integration of  $I_{AB}$,
apart from the prefactor $\omega^8 a^5 t/(8\pi)$, is shown by the dots.
The solid line shows that the asymptotic value of 
$\phi(v)\sim-v^4$ leads
to the power-law behavior $\mathcal{N}/(\omega a)^4$ with
$\mathcal{N}\approx-27$.}
\label{fig4}
\end{minipage}\hfill
\begin{minipage}{.45\textwidth}
\centering
\includegraphics[width=5cm]{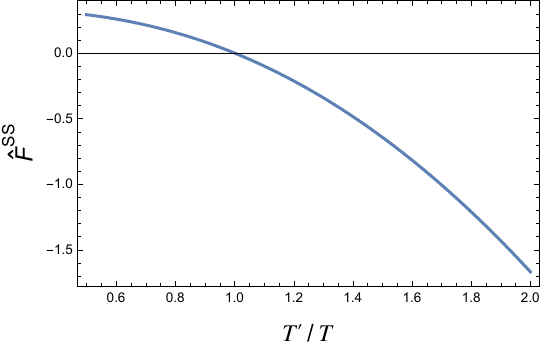}
\caption{Dimensionless force $\hat F$ as a function of the temperature
of the spherical shell relative to the room-temperature background.}
\vskip1cm
\label{fig5}

\end{minipage}
\end{figure}

{\it Example 2: Thin spherical shell.}
Consider next a thin spherical shell of radius $a$ and thickness $t$.  Suppose
the upper and lower hemispheres have different uniform susceptibilities
\begin{equation}
\chi(\mathbf{r};\omega)=\left\{\begin{array}{cc}
0<\theta<\frac{\pi}2:& \chi_A^{\vphantom{q}}(\omega),\\
\frac{\pi}2<\theta<\pi:&\chi_B^{\vphantom{q}}(\omega).\end{array}\right.
\end{equation}
In this case the geometric integral $I_{AB}$ is shown in Fig.~\ref{fig4}.

The force on the thin spherical shell is
\begin{equation}
F=-\frac{\omega_p^2 t^2 a}{16\pi^2}\chi_A^{\vphantom{q}}\nu^3 \mathcal{N}
\hat{F}\approx1 \times 10^{-12}\chi^{\vphantom{q}}_A  \hat{F}\,\mbox{N} \label{forcesph}
\end{equation}
for gold, with  the thickness of the shell being $t=2/\omega_p$ 
(minimum value of skin depth), and the radius of the shell 
being $a=1$ cm.
The dimensionless force $\hat F$ is shown in Fig.~\ref{fig5}.

\begin{figure}
\centering
\begin{minipage}{.45\textwidth}
\centering
\includegraphics[width=6cm]{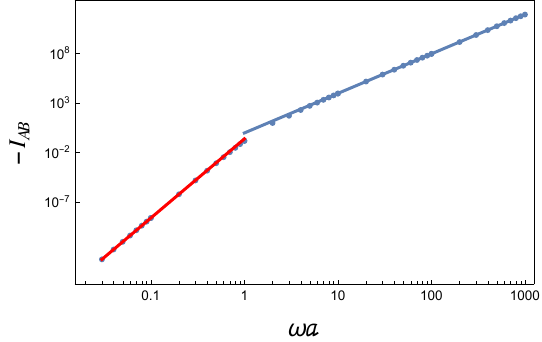}
\caption{The geometric integral for a Janus ball, scaled by $8\pi a$, 
is given by the dots, as a function of $\omega a$. The straight lines
 show the large $\sim(\omega a)^4$ and small $\sim( \omega a)^8$ behaviors.}
\vskip.2cm
\label{fig6}
\end{minipage}\hfill
\begin{minipage}{.5\textwidth}
\centering
\includegraphics[width=6cm]{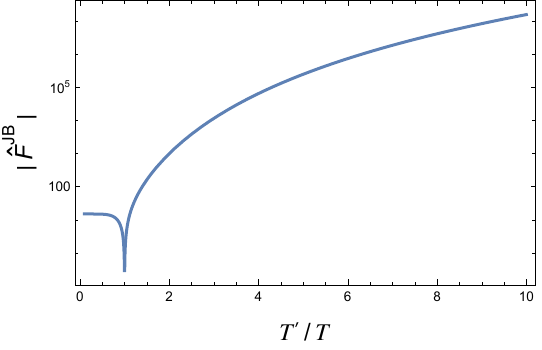}
\caption{Semilog plot of the magnitude of the force on a Janus ball
as a function of the temperature of the ball, relative to that of the
room-temperature background. The force is negative if the ball is hotter
than the blackbody background.}
\label{fig7}
\end{minipage}
\end{figure}

{\it Example 3. Janus ball.}
We now consider a ball, of radius $a$,
each half-ball, $A$ and $B$, being composed of a different
homogeneous material.  Then the
geometric integral 
for such  a ball is
shown in Fig.~\ref{fig6}.

The spontaneous force on a small Janus ball
of radius $1\, \mu$m, with $A$ inert and $B$ gold, is 
\begin{equation}
F^{\rm JB}=\frac1{27\pi}\chi^{\vphantom{q}}_A\omega_p^2(\nu a)^7 \hat{F}\approx
4 \times 10^{-18} \chi^{\vphantom{q}}_A\hat{F} \,\mbox{N}.\label{JBpre}
\end{equation}
 The dimensionless force is shown in 
Fig.~\ref{fig7} for $T=300$ K.
The result may be compared with the 
nonperturbative numerical results of Ref.~\citen{Reid};
the numbers are similar, but the scaling with $a$ is different.

\begin{figure}
\begin{center}
\includegraphics[width=5cm]{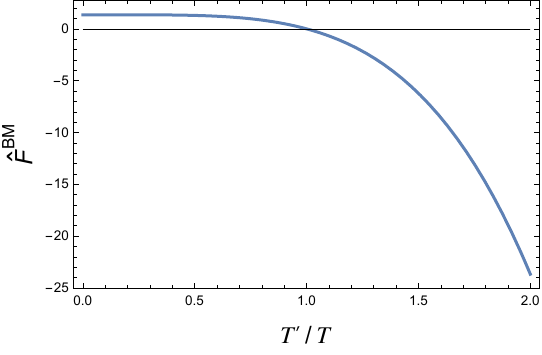}
    \end{center}
\caption{Dimensionless force on a blackbody-metal plate.}
\label{fig8}
\end{figure}

{\it Example 4: Planar structure.} 
We consider a thin planar object, which consists of blackbody material on
the upper side, $A$, and a Drude metal on the lower, $B$.  
By a blackbody material we mean
one whose surface susceptibility,
$\tilde{\chi}=\frac{V}S\chi$, $V$ and $S$ being the volume and the surface area
of the body,  is
\begin{equation}
\tilde{\chi}=\frac{i}4\frac{1}{\omega+i\epsilon},\quad \epsilon\to+0,
\end{equation}
since this will yield Stefan's law for the radiated power,
$P=S\pi^2(T^4-T^{\prime 4})/60$.
If the thicknesses of the two sides are $t_A$ and $t_B$,
the force is
\begin{equation}
F_z=\frac{S t_B(t_A+t_B)\omega_p^2 \nu^4}{24\pi^2}\hat{F},\label{JPF1}
\end{equation}
where $\hat{F}$ is shown in Fig.~\ref{fig8}.
For $S=1$ cm$^2$, $t_A=t_B=10$ nm, and the metal being
gold, the prefactor in
the force is $4\times 10^{-13}$ N, corresponding to an acceleration of 
$\sim 40$~m/s$^2$.  Comparable results were found in 
Ref.~\citen{Manjavacas}.  In this case, as in the other examples, the force
is always towards the metal side of the object (in our convention,
negative).

\subsection{Friction and thermalization lead to terminal velocity}
Once the body starts to move due to the nonequilibrium quantum force,
it will experience quantum friction, nonrelativistically given by
the Einstein-Hopf formula,
\begin{equation}
F_f=-\frac{v\beta}{12\pi^2}\int_0^\infty d\omega\,\omega^5
\Im\alpha(\omega)\frac1{\sinh^2\beta\omega/2},
\end{equation}
and, therefore, it will reach a terminal velocity:
\begin{equation}
v(t)=v^{\vphantom{q}}_T\left(1-e^{-t/t_0}\right).
\end{equation}
For example, a  needle having a 1 cm half-length and 10 nm radius, 
would have a  terminal
velocity  of order 5 m/s.  But the time scale, $t_0 \sim 15$ yrs, is 
very long.

More important is thermal cooling.
Unless there is some mechanism supplied to maintain the temperature difference, the body will eventually cool to the temperature of the
environment.
From Newton's law, the terminal velocity is
\begin{equation}
v^{\vphantom{q}}_T=\frac1m \int_0^\infty dt\, F(T'(t), T).\label{tervelcool}
\end{equation}
Here, the cooling rate is given by
\begin{equation}
    \frac{dQ}{dt}=C_V(T')\frac{dT'}{dt}=P(T',T),
    \end{equation}
where $C_V(T')$ is the specific heat of the body at temperature $T'$.
For a slowly moving body the power radiated is\cite{eqfii}
\begin{equation}
P(T',T)= \frac1{3\pi^2}\int_0^\infty d\omega\,\omega^4\Im
\tr\mbox{\boldmath{$\alpha$}}(\omega)\left[
\frac1{e^{\beta\omega}-1}-\frac1{e^{\beta'\omega}-1}\right].
\end{equation}

In terms of $u=T'/T$, the time taken for the body to cool from temperature
$T_0'$ to temperature $T_1'$,
$T_0'>T_1'>T$, is (for temperatures well above the Debye temperature)
\begin{equation}
t_1=\int_{T_0'}^{T'_1} dT'\frac{C_V(T')}{P(T',T)}, \quad \mbox{or}\quad
   \frac{t_1}{t_c}=\int_{u_0}^{u_1} du \frac1{p(u;T)}, 
\end{equation}
where the dimensionless power radiated is modeled by\cite{nr}
\begin{equation}
p(u;T)=\int_0^\infty dx\frac{x^3}{x^2+1}\left(\frac1{e^{\beta\nu x}-1}-
\frac1{e^{\beta\nu x/u}-1}\right).\label{p}
\end{equation}
Therefore,  the terminal velocity is
\begin{equation}
    v^{\vphantom{q}}_T=\frac{t_c}m \int_{u_0}^1 du \frac{ F(u;T)}{p(u;T)}.
\end{equation}
The time scale for a weak susceptibility model of the  metal portion of
an object is
\begin{equation}
    t_c=\frac{3\pi^2 n T}{\nu^3\omega_p^2}\sim 10^{-4} \,\mbox{s},
    \quad n= \,\mbox{number}\,\mbox{density},
\end{equation}
For a Janus ball, of radius 100 nm,
half made of gold and half dielectric,  
with an initial temperature twice that of the background,
the terminal velocity is only about 0.1 nm/s.  For a 
spherical shell model with radius 1 cm and skin-depth thickness,
$v_T$ would be about the same.
So, observation might  prove challenging.

\section{Spontaneous Torque on an Inhomogeneous Chiral Body}

In Sec.~\ref{sec2}, 
we calculated the torque in first order, which required the body
be composed of nonreciprocal material, which usually necessitates an external
field be applied.  In second order, a torque can arise on a reciprocal
body, but again only if the
body is {\it inhomogeneous}.  It must further be {\it chiral}, in that its mirror reflection cannot be turned into the original object by translations
or rotations.

For a body with isotropic but inhomogeneous susceptibility there is only
an external torque, calculated in second order:
\begin{eqnarray}
\tau_i&=&\int\frac{d\omega}{2\pi}\left(\coth\frac{\beta\omega}2
-\coth\frac{\beta'\omega}2\right)\epsilon_{ijk}
\int(d\mathbf{r})(d\mathbf{r'})X(\mathbf{r,r'};\omega)\nonumber\\
&&\qquad\times\Im\Gamma_{lm}(\mathbf{r-r'};\omega)r_j\nabla_k
\Im\Gamma_{ml}(\mathbf{r'-r};\omega).\label{torque1}
\end{eqnarray}
Again the susceptibility product $X$ [Eq.~(\ref{X})]
vanishes if the body is homogeneous.
If the body consists of two homogeneous parts, $A$ and $B$, $X$ is replaced
by $X_{AB}$ defined in Eq.~(\ref{XAB}).
Using Eq.~(\ref{gradgg}), we find
the torque in this situation is
\begin{subequations}
\begin{equation}
\bm\tau=\frac1{2\pi^2}\int_0^\infty \frac{d\omega}{2\pi}X_{AB}(\omega)
\left(\frac1{e^{\beta\omega}-1}-\frac1{e^{\beta'\omega}-1}\right)
\mathbf{J}_{AB}(\omega),
\end{equation}
in terms of  the geometric factor
\begin{equation}
\mathbf{J}_{AB}(\omega)=-\int_A(d\mathbf{r})\int_B(d\mathbf{r'})
\frac{\mathbf{r\times r'}}{|\mathbf{r-r'}|^8}\phi(\omega|(\mathbf{r-r'}|).
\label{gf}
\end{equation}
\end{subequations}
That the integral is convergent is evident from the behavior of $\phi$ 
given in Eq.~(\ref{limits}).

\begin{figure}
\centering
\begin{minipage}{.45\textwidth}

\centering
\includegraphics[width=\textwidth, trim = 1cm 19cm 11cm 1cm, clip]{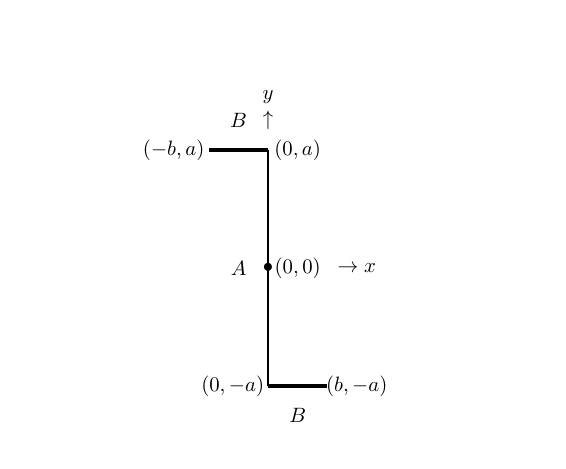}
\caption{
An inhomogeneous wire of small cross section bent in the shape of a dual Allen 
wrench.
The end pieces (``tags'')
 $B$ are taken to be dispersionless dielectric, while the central
 piece $A$ is a Drude-type metal.  The Cartesian coordinates of the
various junctions are shown, as is the center of mass.}
\label{figdaw}
\end{minipage}\hfill
\begin{minipage}{.5\textwidth}
\centering
    \includegraphics[width=\textwidth, trim = 1cm 19cm 11cm 1cm, clip]{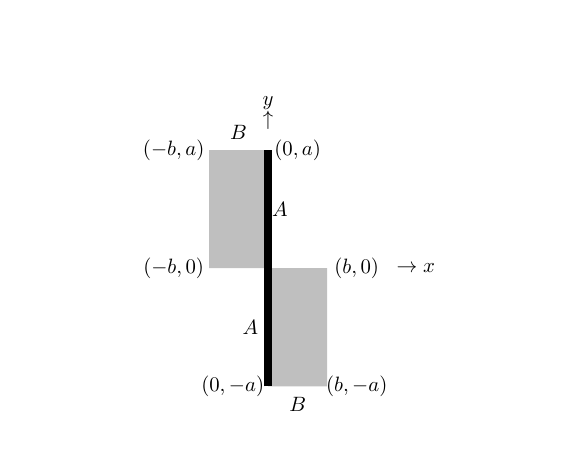}
\caption{Planar inhomogeneous object, obtained by replacing the tags by 
flags.}
\vskip1.8cm
\label{figdf}

\end{minipage}
\end{figure}
\begin{figure}
\centering
\begin{minipage}{.45\textwidth}
\centering
\includegraphics[width=5cm]{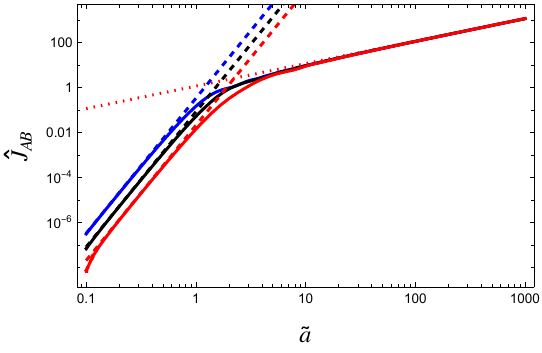}
\caption{Scaled geometric factor as a function of the length of the central
wire, $\tilde{a}=a\omega$,  for different values of $b=a/2$, $a$, or $2a$ 
from bottom to top. These are compared with the 
asymptotes for  large- (dotted) and 
small-  (dashed) arguments.}
\label{figjab} 

\end{minipage}\hfill
\begin{minipage}{.45\textwidth}

\centering
\includegraphics[width=5cm]{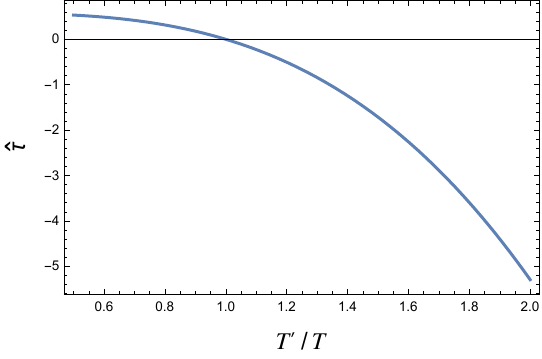}
\caption{Torque, apart from the prefactor, on a
large Allen wrench, as a function
of its temperature relative to the room-temperature background.}
\vskip1cm
\label{figlaw}

\end{minipage}
\end{figure}

{\it Examples: Dual Allen wrench and dual flags.}
A simple example of an inhomogeneous chiral object is shown in 
Fig.~\ref{figdaw}, which illustrates what we might refer to as a dual Allen
wrench.
This
object will experience a quantum vacuum torque, but not a net force, because
it is reflection invariant in the origin, $\mathbf{r\to-r}$.
The geometric factor for this object, which points in the direction
perpendicular to the plane of the object,  is
\begin{equation}
J_{AB}(\omega)=2 S_AS_B\omega^8
\int_{-a}^a dy \int_0^b dx \, x y \frac{\phi(v)}{v^8},
\quad v=\omega\sqrt{x^2+(a+y)^2}.
\end{equation}
$J_{AB}$, apart from a prefactor of $2\omega^4
S_A S_B$, where $S_i$ is the cross-sectional area of the $i$th wire, is shown 
in the Fig.~\ref{figjab} in terms of  $\tilde{a}=\omega a$. 

The large $\tilde{a}$ 
behavior is easily understood: the interactions between the
 parts are local, so increasing $b$ beyond a certain point does not increase 
the torque.  The local forces on $A$ due to the tags $B$
are also saturated as $a$ increases, but 
the lever arm increases linearly. 
The asymptotic value of $J_{AB}$, apart from the same prefactor, is
$\hat{J}_{AB}\sim 11\pi\omega a/30$  for $\omega a\gg1$.
Around room temperature, the transition from large to small occurs at 
$a$, $b\sim 10$ $\mu$m.
The torque on a large Allen wrench is
\begin{equation}
\tau=\frac{11}{60 \pi^2} S_AS_B a \nu^4\omega_p^2
\chi^{\vphantom{q}}_B\hat{\tau},\quad \hat{\tau}=\int_0^\infty dx
\frac{x^4}{x^2+1}\left(\frac1{e^{\beta\nu x}-1}-\frac1{e^{\beta'\nu x}-1}
\right).\label{hitau}
\end{equation}
For a gold wire of circular radius 50 nm, and length $a=1$ cm,
the prefactor, $\tau_0$,  is $7 \times 10^{-22}$~N\,m.
The integral,  $\hat\tau$, is shown in Fig.~\ref{figlaw}.
Such a torque might well be observable, although the terminal
angular velocity
due to cooling is small, $\omega^{\vphantom{q}}_T\sim 10^{-9}$ s$^{-1}$,
if the initial temperature $T'=2T$.

\begin{figure}
\begin{center}
\includegraphics[height=4cm]{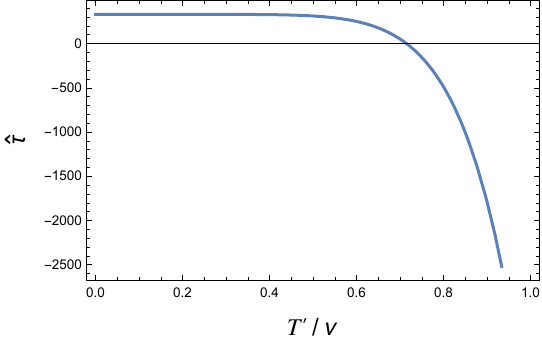}
\end{center}
\caption{Torque, $\hat\tau$, on a small Allen wrench.}
\label{figsaw}
\end{figure}

 The situation might be much more favorable for a small object.  If 
    $\omega a$, $\omega b \ll 1$, the geometric integral is, of course,
    much smaller,
        $\hat{J}_{AB}= 56\omega^6 a^4 b^2/675$.
    Now, the corresponding torque is 
    \begin{equation}
        \tau=\frac{28}{675\pi^3}\chi_B \nu^9\omega_p^2S_AS_B a^4b^2\hat{\tau}.
    \end{equation}
where $\hat\tau$ is shown in Fig.~\ref{figsaw}.
In this case, the terminal angular velocity is much larger, because
the moment of inertia of the object is much smaller as well,
\begin{equation}
I=\rho_A S_A\frac23a^3+\rho_BS_B2b\left(a^2+\frac13 b^2\right).
\end{equation}
so the resulting terminal angular velocity due to thermal cooling is
\begin{equation}
    \omega_T=\frac{t_c\tau_0}I \hat{\omega}_T, \quad \hat{\omega}_T=
    \int_{T'_0/T}^1 du\frac{\hat{\tau}(u;T)}{p(u;T)}
\end{equation}
where the prefactor $t_c\tau_0/I \sim 2\times 10^{-7}$  s$^{-1}$ for a 1 $\mu$m
 object, and $p$ is given by Eq.~(\ref{p}).
Because the dimensionless torque is large, so is
$\hat{\omega}_T\sim 20,000$, if the initial temperature of the object
is twice that of the room temperature environment, leading to a large terninal
angular velocity,
    $\omega_T\sim 4\times 10^{-3}\, \mbox{s}^{-1}$,
which should be quite observable.

Further enhancement of the terminal angular velocity, by about a factor of 
$10^5$ for a large (1 cm) object
or by about a factor of 10 for a small (1 $\mu$m)
object, can be achieved by increasing
the size of the tags, so we have a dual flag shown in Fig.~\ref{figdf}.
Terminal angular velocities for both large and small dual flags
 would seem to be easily accessible to experimental
observation.

\section{Conclusions}
\begin{itemlist}
\item In first order in electric susceptibility, a vacuum torque, but no
force, can arise for a body made of {\it nonreciprocal\/} material.
\item In second order, a vacuum force  can arise only if the body is 
{\it inhomogeneous}, but no exotic material properties are required.
\item A vacuum torque can also arise for ordinary chiral 
bodies in second order, 
but again only if the body is also {\it inhomogeneous}.  This is in contrast 
to the findings of Ref.~\citen{Reid}.
\item We consider several examples of bodies which  exhibit 
possibly observable vacuum forces and torques, although cooling to equilibrium
with the vacuum environment may make it  challenging
to observe the resulting
linear and angular terminal velocities, unless 
the temperature imbalance is maintained.  
The terminal angular velocity for a dual flag
 appears measurable.
\item For dielectric-metal bodies, the force (and torque) is toward the metal 
side,
due to the low emissivity and a high  reflectivity of the metal. 
\item One has to go to third order to see forces and torques on bodies made
of homogeneous material. This will be discussed in future work.

\end{itemlist}

\section*{Acknowledgments}

This work was supported in part by a grant from the US National Science
Foundation, grant number 2008417.
We thank Steve Fulling,
Xin Guo, and  Prachi Parashar
for collaborative assistance.  This paper
reflects solely the authors' personal opinions and does not represent
the opinions of the authors' employers, present and past, in any
way. 


\end{document}